\documentclass[12pt]{article}

\usepackage{graphics}
\usepackage{latexsym, pstricks}

\usepackage{amsmath}

\font\titulo=cmbx10 scaled\magstep1 
scaled\magstep1

\def\section#1{\vskip 1.5truepc plus 0.1truepc minus 0.1truepc
    \goodbreak \leftline{\titulo#1} \nobreak \vskip 0.1truepc
    \indent}
\def\frc#1#2{\leavevmode\kern.1em
    \raise.5ex\hbox{\the\scriptfont0 $ #1 $}\kern-.1em
    /\kern-.15em\lower.25ex\hbox{\the\scriptfont0 $ #2 $}}

\def\IZ{{\rm Z}\llap{\vrule height7.1pt width1pt
     depth-.4pt\phantom t}} 

\newbox\pmbbox
 \def\pmb#1{{\setbox\pmbbox=\hbox{$#1$}%
\copy\pmbbox\kern-\wd\pmbbox\kern.3pt\raise.3pt\copy\pmbbox\kern-\wd\pmbbox
\kern.3pt\box\pmbbox}}

\font\cmss=cmss10 \font\cmsss=cmss10 at 7pt

\def\IZ{\relax\ifmmode\mathchoice
{\hbox{\cmss Z\kern-.4em Z}}{\hbox{\cmss Z\kern-.4em Z}}
{\lower.9pt\hbox{\cmsss Z\kern-.4em Z}} {\lower1.2pt\hbox{\cmsss
Z\kern-.4em Z}}\else{\cmss Z\kern-.4em Z}\fi}

\font\cmss=cmss10 \font\cmsss=cmss10 at 7pt
\def\IS{\relax\ifmmode\mathchoice
{\hbox{\cmss S\kern-.4em S}}{\hbox{\cmss S\kern-.4em S}}
{\lower.9pt\hbox{\cmsss S\kern-.4em S}} {\lower1.2pt\hbox{\cmsss
S\kern-.4em S}}\else{\cmss S\kern-.4em S}\fi}

\begin{document}

\centerline{\titulo On the Robustness of \textit{NK}-Kauffman
Networks Against}

\centerline{\titulo Changes in their Connections and Boolean
Functions}

\vskip 1.2pc \centerline{Federico Zertuche}

\vskip 1.2pc \centerline{Instituto de Matem\'aticas, Unidad
Cuernavaca} \centerline{Universidad Nacional Aut\'onoma de
M\'exico} \centerline{A.P. 273-3, 62251 Cuernavaca, Morelos.,
M\'exico.} \centerline{\tt zertuche@matcuer.unam.mx}

\vskip 3pc {\bf \centerline {Abstract}}

\textit{NK}-Kauffman networks $ {\cal L}^N_K $ are a subset of the Boolean functions on $ N $ Boolean variables to themselves, $ \Lambda_N = \left\{ \xi: \IZ_2^N \to \IZ_2^N \right\} $. To each \textit{NK}-Kauffman network it is possible to assign a unique Boolean function on $N$ variables through the function $ \Psi: {\cal L}^N_K \rightarrow \Lambda_N $. The probability $ {\cal P}_K $ that $ \Psi \left( f \right) = \Psi \left( f '\right) $, when $ f' $ is obtained through $ f $ by a change of one of its $K$-Boolean functions ($ b_K: \IZ_2^K \to \IZ_2 $), and/or connections; is calculated. The leading term of the asymptotic expansion of $ {\cal P}_K $, for $ N \gg 1 $, turns out to depend on: the probability to extract the \textit{tautology} and \textit{contradiction} Boolean functions, and in the average value of the distribution of probability of the Boolean functions; the other terms decay as $ {\cal O} \left( 1 / N \right) $. In order to accomplish this, a classification of the Boolean functions in terms of what I have called their {\it irreducible degree of connectivity} is established. The mathematical findings are discussed in the biological context where, $ \Psi $ is used to model the genotype-phenotype map.

\vskip 2pc

\noindent {\bf Short title:} {\it Robustness Against Mutations.}

\vskip 1pc \noindent {\bf Keywords:} {\it Cellular automata,
irreducible connectivity, binary functions, functional graphs,
redundant genetic material, genetic robustness}.

\vskip 1pc \noindent {\bf PACS numbers:} 05.65.+b, 87.10.+e,
87.14.Gg, 89.75.Fb

\newpage

\baselineskip = 12.4pt

\section{1. Introduction}

\textit{NK}-Kauffman networks are useful models for the study the genotype-phenotype map $ \Psi $; which is their main application in this work~${}^{1,2}$. An \textit{NK}-Kauffman network consists of $ N $ Boolean variables $ S_i (t) \in \IZ_2 $ ($ i = 1, \dots, N $), that evolve deterministically in discretized time $ t = 0, 1, 2, \dots $ according to Boolean functions on $ K $ ($ 0 \leq K \leq N $) of these variables at the previous time $ t-1 $. For every site $ i $, a $K$-Boolean function $ f_i: \IZ_2^K \to \IZ_2 $ is chosen randomly and independently with a bias probability $p$ ($ 0 < p < 1 $), that $ f_i = 1 $ for each of its possible $ 2^K $ arguments. Also, for every site $ i $, $ K $ inputs (the connections) are randomly selected from a uniform distribution, among the $ N $ Boolean variables of the network, without repetition. Once the $ K $ inputs and the functions $ f_i $ have been selected, a Boolean deterministic dynamical system; known as a \textit{NK}-Kauffman network has been constructed. The network evolves deterministically, and synchronously in time, according to the rules
$$
S_i (t+1) = f_i \left( S_{i_1}(t), S_{i_2}(t), \dots, S_{i_K}(t)
\right), \ \ i = 1, \dots, N, \eqno(1)
$$
where $ i_m \not= i_n $, for all $ m, n = 1, 2, \dots, K $, with $ m \not= n $; because all the inputs are different. An \textit{NK}-Kauffman network is then a map of the form
$$
f: \IZ_2^N \longrightarrow \IZ_2^N .
$$
Let us denote by  $ {\cal L}^N_K $ the set of \textit{NK}-Kauffman networks that might be built up, by this procedure, for given $ N $ and $ K $. They constitute a subset of the set of all possible Boolean functions on $N$-Boolean variables to themselves
$$
\Lambda_N = \left\{ \xi: \IZ_2^N \to \IZ_2^N \right\}.
$$
In Ref.~1, a study of the injective properties of the map
$$
\Psi : {\cal L}^N_K \rightarrow \Lambda_N \eqno(2)
$$
was pursuit for the case  $ p = 1 / 2 $; where the Boolean functions are extracted from a uniform distribution. Using the fact that $ \Lambda_N \cong {\cal G}_{2^N} $, where $ {\cal G}_{2^N} $ is the set of functional graphs on $ 2^N $ points~${}^{3}$; the average number $ \vartheta \left( N, K \right) $ of elements in $ {\cal L}^N_K $ that $ \Psi $ maps into the same Boolean function was calculated~${}^{1}$. The results showed that for, $ K \sim {\cal O} \left( 1 \right) $ when $ N \gg 1 $, there exists a critical average connectivity
$$
K_c \approx \log_2 \log_2 \left( {2 N \over \ln 2} \right) + {\cal
O} \left( {1 \over N \ln N } \right); \eqno(3)
$$
such that $ \vartheta \left( N, K \right) \approx e^{\varphi \, N} \gg 1 $ ($ \varphi > 0 $) or $ \vartheta \left( N,K \right) \approx 1 $, depending on whether $ K < K_c $ or $ K > K_c $, respectively.

In genetics, \textit{NK}-Kauffman networks are used as models of the genotype-phenotype map, represented by $ \Psi $~${}^{1,2}$: The genotype consists in a particular wiring, and selection of the Boolean functions $ f_i $ in (1), which give rise to the \textit{NK}-Kauffman network; while the phenotype is represented by their attractors in $ \Psi \left( {\cal L}^N_K \right) \subseteq \Lambda_N \cong {\cal G}_{2^N} $~${}^{4-6}$. The $ K $ connections represent the average number of epistatic interactions among the genes, and the Boolean variables $ S_i $; the expression ``1'' or inhibition ``0'' of the {\it i}-th gene.

A well established fact in the theory of natural selection is the so-called robustness of the phenotype against mutations in the genotype~${}^{1,2,7,8}$. At the level of the genotype, random mutations (by radiation in the environment) and recombination by mating, constitute the driving mechanism of the {\it Evolution Theory}. Experiments in laboratory controlling the amount of radiation have shown that; while the change in the phenotype vary from species to species, more than $ 50 \% $ of the changes have no effect at all in the phenotype~${}^{8-11}$. In Ref.~1 it was shown that the signature of genetic robustness can be seen in the injective properties of $ \Psi $, with a many-to-one map representing a robust phase. This happens if $ K < K_c $, with the value of  $ N $ to be substituted on (3), determined by the number of genes that living organisms have. This number varies from $ 6 \times 10^3 $ for yeast to less than $ 4 \times 10^4 $ in {\it H. sapiens}~${}^{9}$. Substitution in (3) gives in both cases that $ K \leq 3 $~${}^{1}$.

In this article it is calculated; for a general bias $ p $, the probability $ {\cal P}_K $ that two elements $ f, f' \in {\cal L}^N_K $, such that $ f' $ is obtained from $ f $ by a mutation, give rise to the the same phenotype, {\it i.e.} $ \Psi \left( f \right) = \Psi \left( f' \right) $. For a mutation, it is intended a change in a Boolean function $ f_i $, and/or its connections. The results impose restrictions in the values that $ K $ and $ p $ should have, in order that $ {\cal P}_K \geq 1/2 $, in accordance with the experiments.

The article is organized as follows: In Sec.~2, I set up a mathematical formalism that allows to write (1) in a more suitable way for calculations. In Sec.~3, the expressions of the different probabilities involved in the calculation of $ {\cal P}_K $ are established. In Sec.~4, I introduce a new classification of Boolean functions according to its real dependence on their arguments; which I call its {\it degree of irreducibility}. This classification is used in Sec.~5 to calculate the invariance of Boolean functions under changes of their connections and so; calculate $ {\cal P}_K $. In Sec.~6 the conclusions are set up. In the appendix, two errata of Ref.~1 are corrected, and  it is shown that they do not alter the asymptotic results of Ref.~1. So, the biological implications there stated remain correct.

\bigskip


\section{2. Mathematical Framework}

Now we introduce a mathematical formalism with the scope of write (1) in a more suitable notation for counting. All additions between elements of $ \IZ_2 $ and its cartesian products are modulo $ 2 $.

Let $ {\cal M}_N = \left\{ 1, 2, \dots, N \right\} $ denote the set of the
first $ N $ natural numbers. A $K$-connection set $ C_K $, is any subset of $
{\cal M}_N $ with cardinality $ K $. Since there are $ {N \choose K} $ possible
$K$-connection sets; we count them in some unspecified order, and denote them by
$$
C_K^{(\alpha)} = \left\{ i_1, i_2, \dots, i_K \right\} \subseteq
{\cal M}_N, \ {\rm with} \ \alpha=1,\dots,{N \choose K} \ , \eqno(4)
$$
where, without lost of generality; $ i_1 < i_2 < \dots < i_K $, with $ 1 \leq i_m \leq N $ ($ 1 \leq m \leq K $). We also denote as
$$
\Gamma^N_K = \left\{ C_K^{(\alpha)} \right\}_{\alpha =1}^{{N
\choose K}} \eqno(5)
$$
the set of all $K$-connections. To each $K$-connection set $
C_K^{(\alpha)} $ it is possible to associate a $K$-connection map
$$
 C_K^{* (\alpha)}: \IZ_2^N \longrightarrow \ \IZ_2^K ,
$$
defined by
$$
C_K^{* (\alpha)} \left( {\bf S} \right) = C_K^{* (\alpha)} \left(
S_1, \dots, S_N \right) = \left( S_{i_1}, \dots, S_{i_K} \right) \ \ \forall \ {\bf S} \in \IZ_2^N.
$$
Any map
$$
b_K: \IZ_2^K \to \IZ_2, \eqno(6)
$$
defines a $K$-Boolean function. Since $ \# \IZ_2^K = 2^K $; $ b_K
$ is completely determined by its $K$-truth table $T_K$, given by
$$
T_K = \left[ A_K \ {\bf b}_K \right],
$$
where $ A_K $ is a $ 2^K \times K $ binary matrix, and $ {\bf b}_K
$ is a $ 2^K $ dimensional column-vector, such that:

The $s$-th row ($ 1 \leq s \leq  2^K $) of $ A_K $ encodes the binary decomposition of $s$, and represents each one of the possible $ 2^K $ arguments of $ b_K $ in (6). So, $ A_K $ satisfies~${}^{1}$
$$
s = 1 + \sum_{i=1}^K A_K \left( s, i \right)  2^{i-1}.
$$
And
$$
{\bf b}_K = \left[ \sigma_1, \sigma_2, \dots, \sigma_{2^K}
\right], \eqno(7)
$$
where $ \sigma_s \in \IZ_2 $ ($ 1 \leq s \leq 2^K $), represents the images of (6).

There are as much as $2^{2^K}$ $K$-truth tables $ T_K $ corresponding to the total possible vectors (7). $K$-Boolean functions can be classified according to Wolfram's notation by their decimal number $ \mu $ given by~${}^{1,12}$
$$
\mu = 1 + \sum_{s=1}^{2^K} 2^{s-1} \sigma_s. \eqno(8)
$$
Let us add a superscript $ (\mu) $ to a $K$-Boolean function, or to its truth table, whenever we want to identify them. So, $ b^{(\mu)}_K $ and $ T_K^{(\mu)} $ refer to the $\mu$-th $K$-Boolean function and its truth table respectively. Within this notation, the set of all $K$-Boolean functions $ \Xi_K $ is expressed as
$$
\Xi_K = \left\{ b^{(\mu)}_K \right\}_{\mu = 1}^{2^{2^K}}.
$$
Of particular importance are the {\it tautology} $ b_K^{(\tau)}
\equiv b^{(2^{2^K})}_K $ and {\it contradiction} $ b_K^{(\kappa)}
\equiv b^{(1)}_K $ $K$-Boolean functions; with images:
$$
{\bf b}_K^{(\tau)} = \left[ 1, 1, \dots, 1 \right]  \eqno(9a)
$$
and
$$
{\bf b}_K^{(\kappa)} = \left[ 0, 0, \dots, 0 \right].  \eqno(9b)
$$

Table~1 gives an example of the $ A_2 $ matrix (representing
the four possible entries of $ S_1 $ and $ S_2 $) with the sixteen
possible $2$-Boolean functions listed according to their decimal
number (8).

\vfill
\eject

\footnotesize
\begin{center}
\begin{tabular}{|c|c|c|c|c|c|c|c|c|c|c|c|c|c|c|c|c|c|c|}
\hline  $ S_1 $ & $ S_2 $ & $ \mu \ \mapsto $ & {\bf 1} & {\bf 2}
& {\bf 3} & {\bf 4} & {\bf 5} & {\bf 6} & {\bf 7} & {\bf 8} & {\bf
9} & {\bf 10} & {\bf
11} & {\bf 12} & {\bf 13} & {\bf 14} & {\bf 15} & {\bf 16} \\
\hline 0 & 0 &  $ \sigma_1 \mapsto $ & 0 & 1 & 0 & 1 & 0 & 1 & 0 & 1 & 0 & 1 & 0 & 1 & 0 & 1 & 0 & 1 \\
\hline 1 & 0 &  $ \sigma_2 \mapsto $ & 0 & 0 & 1 & 1 & 0 & 0 & 1 & 1 & 0 & 0 & 1 & 1 & 0 & 0 & 1 & 1 \\
\hline 0 & 1 &  $ \sigma_3 \mapsto $ & 0 & 0 & 0 & 0 & 1 & 1 & 1 & 1 & 0 & 0 & 0 & 0 & 1 & 1 & 1 & 1 \\
\hline 1 & 1 &  $ \sigma_4 \mapsto $ & 0 & 0 & 0 & 0 & 0 & 0 & 0 & 0 & 1 & 1 & 1 & 1 & 1 & 1 & 1 & 1 \\
\hline
\end{tabular}
\end{center}
\normalsize \baselineskip = 12.4pt

\centerline{\textbf{Table~1.} The $ A_2 $ matrix, with the sixteen $2$-Boolean functions.}

\

Within the preceding notation, the dynamical rule (1), now may be rewritten
as
$$
S_i \left( t + 1 \right) = b_K^{(\mu_i)} \circ C_K^{* (\alpha_i)}
\left( {\bf S} \left( t \right) \right), \ \ i = 1, \dots, N;
\eqno(10)
$$
where, some of the indexes $ \alpha_i $ and $ \mu_i $ may be
equal for different values of $i$, and $ {\bf S} \left( t \right) \in \IZ_2^N $.

\bigskip


\section{3. The Invariance of \textit{NK}-Kauffman Networks}

Now we are interested in calculate the probability $ {\cal P}_K $, that (10) remains invariant under a change of a connection $ C^{(\alpha)}_K $ and/or a $K$-Boolean function $ b_K^{(\mu)} $. So, we must study the number of ways in which this could happen; {\it i.e.} what conditions should prevail in order that for some $ i $,
$$
b_K^{(\mu_i)} \circ C_K^{* (\alpha_i)} \left( {\bf S} \right) +
b_K^{(\nu_i)} \circ C_K^{* (\beta_i)} \left( {\bf S} \right) \ = \ 0 \ \ \
\forall \ {\bf S} \in \IZ_2^N, \eqno(11a)
$$
for $ \alpha_i \neq \beta_i $ and/or $ \mu_i \neq \nu_i $. Let us use a shorthand notation and skip to write the indexes $ \alpha_i $, and $ \mu_i $. Then (11a) may be  written as
$$
\tilde{b}_K \circ \tilde{C}_K^{*} \left( {\bf S} \right) + b_K \circ C_K^{*} \left( {\bf S} \right)  =  0 \ \ \ \forall \ {\bf S} \in \IZ_2^N,
$$
where, $ \tilde{b}_K =  b_K + \Delta b_K $ and $ \tilde{C}_K^{*} = C_K^{*} + \Delta C_K^{*} $; with $ \Delta b_K  \in \Xi_K $, and $  \Delta C_K^{*} $ a $K$-connection map. Explicit substitution gives
$$
b_K \circ \Delta C_K^{*} \left( {\bf S} \right) + \Delta b_K \circ \tilde{C}_K^* \left( {\bf S} \right)  =  0 \ \ \ \forall \ {\bf S} \in \IZ_2^N. \eqno(11b)
$$
Equation (11b) could be satisfied in three different ways:

\begin{itemize}

\item[{\it i})] Event ${\cal A}$: A change in a $K$-connection $ C_K $ without a change in a $K$-Boolean function $ b_K $. This implies $ \Delta b_K = 0 $ $ \forall \, {\bf S} \in \IZ_2^K $, and from (9b) $ \Rightarrow $ $ b_K \circ \Delta C_K^{*} \left( {\bf S} \right) = b_N^{(\kappa)} $.

\item[{\it ii})] Event ${\cal B}$: A change in a $K$-Boolean function $ b_K $ without a change in a $K$-connection $ C_K $. This implies $ \Delta C_K^{*} = 0 $ $ \forall \, {\bf S} \in \IZ_2^N $. From (9b)  there follows $ \Delta b_K \circ \tilde{C}_K^* \left( {\bf S} \right) = 0 $ $ \forall \, {\bf S} \in \IZ_2^N $ $ \Rightarrow $ $ \Delta b_K = b_K^{(\kappa)} $. So, the $K$-Boolean function must remain unchanged.

\item[{\it iii})] Event ${\cal C}$: A change in a $K$-Boolean function $ b_K $ and a change in a $K$-connection $ C_K $. In this case, both $ \Delta C_K^{*} \not= 0 $ and $ \Delta b_K \not= 0 $; and also (11b) must hold with independency of ${\cal A}$, and ${\cal B}$ events. So, from (9a), it must happen that $ b_K \circ \Delta C_K^{*} = b_N^{(\tau)} $, and $ \Delta b_K \circ \tilde{C}_K^* = b_N^{(\tau)} $.

\end{itemize}

Since the events ${\cal A}$, ${\cal B}$, and ${\cal C}$, are independent, the probability $ {\cal P}_K $ that (11) are satisfied, is given by the combined probabilities $ P \left( {\cal A} \right) $, $ P \left( {\cal B} \right) $, and $ P \left( {\cal C} \right) $ that ${\cal A}$, ${\cal B}$, and ${\cal C}$ happen. So,
\begin{eqnarray}
{\cal P}_K&=&P \left( {\cal A} \right) + P \left( {\cal B} \right) + P \left( {\cal C} \right) - P \left( {\cal A} \right) \, P \left( {\cal B} \right) - P \left( {\cal A} \right) \, P \left( {\cal C} \right) - P \left( {\cal B} \right) \, P \left( {\cal C} \right)
\nonumber\\
&+& P \left( {\cal A} \right) \, P \left( {\cal B} \right) \, P \left( {\cal C} \right). \ \ \ \ \ \ \ \ \ \ \ \ \ \ \ \ \ \ \ \ \ \ \ \ \ \ \ \ \ \ \ \ \ \ \ \ \ \ \ \ \ \ \ \ \ \ \ \ \ \ \ \ \ \ \ \ (12) \nonumber
\end{eqnarray}

For a general bias $ p $ ($ 0 < p < 1 $) that $ \sigma_s = 1 $, for $ 1 \leq s \leq 2^K $ in (7), the probability $ \Pi \left( b_K \right) $ to extract the $K$-Boolean function $ b_K
$ is given by
$$
\Pi \left( b_K \right) = p^\omega \left( 1 - p \right)^{2^K - \,
\omega} \ , \eqno(13a)
$$
where
$$
\omega = \omega \left( b_K \right) = \sum_{s=1}^{2^K} \sigma_s, \eqno(13b)
$$
is the {\it weight} of $ b_K $.

The following considerations are in order:

\begin{itemize}

\item[{\it i})] $ P \left( {\cal A} \right) $ is the probability that the {\it projected function}
    $$
    b_K^{* (\alpha)} \equiv b_K \circ C_K^{*(\alpha)}: \IZ_2^N \rightarrow \IZ_2 \eqno(14)
    $$
    remains invariant under a change of the $K$-connection. To get read of this we must first introduce the concept of irreducibility of Boolean functions, which is going to be done in the next section.

\item[{\it ii})] $ P \left( {\cal B} \right) $ is the average probability that $ b_K $ remains invariant by a mutation, given that $ b_K $ has occurred. Then
    $$
    P \left( {\cal B} \right) = \sum_{b_K \in \, \Xi_K} \Pi^2 \left( b_K \right). $$
    Since there are $ {2^K \choose \omega} $ $K$-Boolean functions with weight $ \omega $, from (13) we obtain
    $$
    P \left( {\cal B} \right) = \sum_{\omega=0}^{2^K} {2^K \choose \omega} \ p^{2 \omega} \ \left( 1 - p \right)^{2^{K+1} - 2 \omega} = \left[ 1 - 2 p \left( 1 - p \right) \right]^{2^K}.
    $$

\item[{\it iii})] $ P \left( {\cal C} \right) $ is the probability of extracting twice the {\it tautology} $N$-Boolean function. From (9a) and (13b) $ \omega \left( b_N^{(\tau)} \right) = 2^N $, so from (13a)
    $$
    P \left( {\cal C} \right) = \Pi^2 \left( b_N^{(\tau)} \right) = p^{2^{N+1}} \ll 1 .
    $$

\end{itemize}

So, we obtain the asymptotic expression for (12)
$$
{\cal P}_K \approx P \left( {\cal A} \right) + \left[ 1 - 2 p \left( 1 - p \right) \right]^{2^K} \left[ 1 - P \left( {\cal A} \right) \right] + {\cal O} \left( p^{2^{N+1}} \right), \eqno(15)
 $$
for $ N \gg 1 $.

\bigskip


\section{4. The Irreducibility of the Boolean Functions}

Not all the $K$-Boolean functions depend completely on their $K$ arguments. For instance, let us consider the $2$-Boolean functions of table~1: Rules {\bf 1} and {\bf 16} ({\it contradiction} and {\it tautology}, respectively) do not depend on either $ S_1 $ or $ S_2 $; while rules {\bf 6} and {\bf 11} ({\it negation} and {\it identity}, respectively) only depend on $S_1$. Due to this fact, let us make the following definitions:

\


\leftline{{\bf Definition 1}}

\noindent A $K$-Boolean function $ b_K $ is {\it reducible} on the $m$-th argument $
S_m $ ($ 1 \leq m \leq K $), if
$$
b_K \left( S_1, \dots, S_m, \dots, S_K \right) = b_K \left( S_1,
\dots, S_m + 1, \dots, S_K \right) \hskip 0.4cm \forall \ {\bf S} \in \IZ_2^K .
$$
Otherwise, the $K$-Boolean function $ b_K $ is {\it irreducible} on the $m$-th argument $ S_m $.

\

\vfill
\eject

\leftline{{\bf Definition 2}}

\noindent A $K$-Boolean function $ b_K $ is {\it irreducible of degree} $ \lambda $ ($ 0 \leq \lambda \leq K $); if it is irreducible on $ \lambda $ arguments and reducible on the remaining $ K - \lambda $ arguments. If $ \lambda = K $, the $K$-Boolean function is {\it irreducible}.

Let us denote by $ {\cal I}_K \left( \lambda \right) $ the set of
irreducible $K$-Boolean functions of degree $ \lambda $. From
definitions 1 \& 2, $ \Xi_K $ may be decomposed uniquely in terms of $ {\cal I}_K \left( \lambda \right) $ by
$$
\Xi_K = \bigcup_{\lambda = 0}^K \ {\cal I}_K \left( \lambda
\right), \eqno(16a)
$$
with
$$
{\cal I}_K \left( \lambda \right) \cap {\cal I}_K \left(
\lambda'\right) = \emptyset \ \ {\rm for} \ \ \lambda \neq \lambda'. \eqno(16b)
$$

The cardinalities $ \beta_K \left( \lambda \right) \equiv \# {\cal I}_K \left( \lambda \right) $ may be calculated recursively, noting that $ \beta_K \left( \lambda \right) $, must be equal to the number of ways to form $ \lambda $ irreducible arguments from $K$ arguments. This amounts to $ {K \choose \lambda} $ times the number of irreducible $\lambda$-Boolean functions $ \beta_\lambda \left( \lambda \right) $; thus
$$
\beta_K \left( \lambda \right) = {K \choose \lambda} \
\beta_\lambda \left( \lambda \right). \eqno(17)
$$
Setting $ K = \lambda $ in (16a) and calculating the cardinalities, follows that
$$
2^{2^\lambda} = \sum_{\nu = 0}^{\lambda - 1} \beta_\lambda \left(
\nu \right) + \beta_\lambda \left( \lambda \right).
$$
Substituting back into (17) the following recursion formulas for the number of irreducible $K$-Boolean functions of degree $ \lambda $ are obtained
$$
\beta_K \left( \lambda \right) = {K \choose \lambda} \ \left[
2^{2^\lambda} - \sum_{\nu = 0}^{\lambda - 1} \beta_\lambda \left(
\nu \right) \right], \eqno(18a)
$$
and
$$
\beta_K \left( 0 \right) = 2.  \eqno(18b)
$$

Note from (9), that $ b_K^{(\tau)} $ and $ b_K^{(\kappa)} $ are irreducible of degree zero. So from (18b),
$$
{\cal I}_K \left( 0 \right) = \left\{ b_K^{(\tau)}, \,
b_K^{(\kappa)} \right\}. \eqno(19)
$$
Some first values for $ \beta_K \left( \lambda \right) $ are:
$$
\beta_K \left( 1 \right) = 2  K  ,
$$
$$
\beta_K \left( 2 \right) = 5  K \left( K - 1 \right) ,
$$
$$
\beta_K \left( 3 \right) = {109 \over 3}  K \left( K - 1 \right) \left( K - 2 \right) ,
$$
$$
\beta_K \left( 4 \right) = {32,297 \over 12}  K \left( K - 1 \right) \left( K - 2 \right) \left( K - 3 \right) ,
$$
etc.

\bigskip

\section{5. The Probability $ P \left( {\cal A} \right) $}

Let us now calculate $ P \left( {\cal A} \right) $ to obtain $ {\cal P}_K $ from (15).
The probability $ P \left( {\cal A} \right) $, that $ b_K^{* (\alpha)} $, defined by (14), remains invariant against a change in $ C_K^{(\alpha)} $, depends in the degree of irreducibility of $ b_K $; {\it i.e.} on which of its $K$ arguments it really depends. To calculate it, let us first calculate the probability $ P \left[ \Delta \ b_K^{* (\alpha)} = 0 | b_K \in {\cal I}_K \left(\lambda \right) \right] $ that, $ b_K^{* (\alpha)} $ remains invariant due to a change in the $K$-connection $ C_K^{(\alpha)} $; given that $ b_K $ is irreducible of degree $ \lambda $.

Let $ b_K \in {\cal I}_K \left(\lambda \right) $ be irreducible in the arguments with indexes
$$
m_1, m_2, \dots, m_\lambda \ , \ {\rm where} \ \ m_1 < m_2 < \dots < m_\lambda
$$
such that $ 1 \leq m_l \leq K $ ($ 1 \leq l \leq \lambda $). Let us also rewrite (4) more explicitly putting the superscript $ (\alpha) $ into its elements; then
$$
C_K^{(\alpha)} = \left\{ i_1^{(\alpha)}, i_2^{(\alpha)}, \dots, i_K^{(\alpha)} \right\} \subseteq {\cal M}_N.
$$
Now, associated to $ b_K^{* (\alpha)} $, we can define its {\it $\lambda$-irreducible connection} by
$$
{\cal J}_\lambda \left( b_K^{* (\alpha)} \right) \equiv \left\{ i_{m_l}^{(\alpha)} \right\}_{l=1}^\lambda \subseteq C_K^{(\alpha)}.
$$
Within this notation the set $ \Theta_K^N \left( b_K^{* (\alpha)} \right) $, of the $K$-connections $ C_K^{(\beta)} $ that leave $ b_K^{* (\alpha)} $ invariant, is given by
$$
\Theta_K^N \left( b_K^{* (\alpha)} \right) = \left\{
C_K^{(\beta)} \in \Gamma_K^N \ | \ i_{m_l}^{(\beta)} = i_{m_l}^{(\alpha)} \ \forall \ l = 1, 2, \dots, \lambda \right\}. \eqno(20)
$$
Then
$$
P \left[ \Delta \, b_K^{* (\alpha)} = 0 | b_K \in {\cal I}_K \left(\lambda
\right) \right] = {\# \Theta_K^N \left( b_K^{* (\alpha)} \right)
\over \# \Gamma_K^N} \ . \eqno(21)
$$
From (5), $ \# \Gamma_K^N = {N \choose K} $. To calculate $ \# \Theta_K^N \left( b_K^{* (\alpha)} \right) $, let us note that the $K$-connections $ C_K^{(\beta)} \in \Theta_K^N \left( b_K^{* (\alpha)} \right) $ have $ \lambda $ elements fixed, the elements of $ {\cal J}_\lambda \left( b_K^{* (\alpha)} \right) $, and $ K - \lambda $ elements free, which are the elements of $ {\cal M}_N \setminus {\cal J}_\lambda \left( b_K^{* (\alpha)} \right) $. Thus, $ \# \Theta_K^N \left( b_K^{* (\alpha)} \right) $ equals the number of subsets of $ {\cal M}_N \setminus {\cal J}_\lambda \left( b_K^{* (\alpha)} \right) $ that can be constructed with $ K - \lambda $ elements. Since
$$
\# \left[ {\cal M}_N \setminus {\cal J}_\lambda \left( b_K^{* (\alpha)} \right) \right] = N - \lambda,
$$
we obtain
$$
\# \Theta_K^N \left( b_K^{* (\alpha)} \right) = {N - \lambda
\choose K - \lambda}. \eqno(22)
$$
That only depends in the degree of irreducibility $ \lambda $ of  $ b_K $ and not in the connection index $ (\alpha) $. Substituting (22) into (21) we obtain
$$
P \left[ \Delta \, b_K^{* (\alpha)} = 0 | b_K \in {\cal I}_K \left(\lambda
\right) \right] = {K! \left( N - \lambda \right)! \over N! \left(
K - \lambda \right)!} \ . \eqno(23a)
$$

Due to (16), $ P \left( {\cal A} \right) $ is given by:
$$
P \left( {\cal A} \right) = \sum_{\lambda = 0}^K P \left[ \Delta \, b_K^{* (\alpha)} = 0 | b_K \in {\cal I}_K \left(\lambda \right) \right] \ P \left[ b_K \in {\cal
I}_K \left(\lambda \right) \right] , \eqno(23b)
$$
where $ P \left[ b_K \in {\cal I}_K \left(\lambda \right) \right]
$ is the probability that $ b_K $ be irreducible of de\-gree $
\lambda $. The value of $ P \left[ b_K \in {\cal I}_K \left(\lambda \right)
\right] $ depends on $ \beta_K \left( \lambda \right) $ [calculated from (18)], as well as on the particular way in which the $K$-Boolean
functions $ b_K $ are extracted.

When $ K \sim {\cal O} \left( 1 \right) $ for $ N \gg 1 $, equations (23) behave asymptotically like
$$
P \left( {\cal A} \right) \approx P \left[ b_K \in {\cal I}_K \left( 0 \right) \right] + {\cal O} \left( {1 \over N} \right).
$$
So from (19), the leading term of $ P \left( {\cal A} \right) $ comes from the probability to extract the {\it tautology} (9a) and {\it contradiction} (9b)
$K$-Boolean functions. We obtain from (13)
$$
P \left( {\cal A} \right) \approx p^{2^K} + \left( 1 - p \right)^{2^K} + {\cal O} \left( {1 \over N} \right).
$$
From (15) the probability that (1) [or equivalently (10)] remains invariant by a change on a $K$-Boolean function and/or its connection; is given by
\begin{eqnarray*}
{\cal P}_K &\approx& p^{2^K} + \left( 1 - p \right)^{2^K} \\
&+& \left[ 1 - 2 p \left( 1 - p \right) \right]^{2^K} \left\{ 1 - \left[ p^{2^K} + \left( 1 - p \right)^{2^K} \right] \right\} + {\cal O} \left( {1 \over N} \right). \ \ \ \ \ (24) \nonumber
\end{eqnarray*}

\bigskip


\section{6. Conclusion}

A classification of $K$-Boolean functions in terms of its irreducible degree of connectivity $ \lambda $ was introduced. This allowed us to uniquely decompose them through (16), and calculate the asymptotic formula (24) for $ {\cal P}_K $; that an \textit{NK}-Kauffman network (1) remains invariant against a change in a $K$-Boolean function and/or its $K$-connection. Figure~1 shows the graphs for $ {\cal P}_K $ {\it vs} $ p $; for different values of the average connectivity $ K $. The graphs attain a minimum and are symmetric at $ p = 1/2 $ (the case of a uniform distribution). For $p$ fixed, $ {\cal P}_K < {\cal P}_{K'} $ for $ K > K' $.

These results are specially important when \textit{NK}-Kauffman network are used to model the genotype-phenotype map (2)~${}^{1,2}$. Experiments to study the robustness of the genetic material have been done by means of induced mutations~${}^{9-11}$. The results varied among the different organisms studied, but it is estimated that in more than $ 50 \% $ of the cases the phenotype appears not to be damaged. In \textit{NK}-Kauffman networks this phenomena is manifest when $ {\cal P}_K > 1/2 $. Figure~1 shows that is possible to be in agreement with the experimental data without a bias ($ p = 1/2 $), provided $ K \leq 1.25 $ for the average connectivity. For the case $ K = 2 $ this happens only for values of $ p $ outside the interval $ [0.21, 0.78] $. There is no surprise that biassed values of $ p $ increment the value of $ {\cal P}_K $ since they tend to increase the amount of {\it tautology} and {\it contradiction} functions (9) through (13).

\bigskip


\section{Acknowledgments}

This work is supported in part by {\bf CONACyT} project number {\bf 059869} and {\bf PAPIIT} project number {\bf IN101309-3}. The author wishes to thank: Martha Takane for fruitful mathematical discussions, Thal\'\i a Figueras for careful reading of the manuscript, Mamed Atakishiyev for computational advice and Pilar L\'opez Rico for accurate services on informatics.

\bigskip


\section{Appendix: Errata in Ref.~1}

All quotations to equations in Ref.~1 are preceded by an ``R'', those introduced here by an ``A'', while all the others refer to equations of the present article.

In Ref.~1 it was wrongly stated that the only Boolean functions that contribute to the number of redundances $ r $ in (R16) are: the {\it tautology}, the {\it contradiction}, the {\it identity} and the {\it negation}. In fact there are contributions from many more functions, their number growing with $K$ for $ K < N $ (in the case $ K = N $ of the {\it random map model} $ r = 0 $; as explained further); according to their classification in terms of its degree of irreducibility defined in Sec.~4 of this article. Furthermore; the contribution to $r$ of the {\it identity} and {\it negation} functions were calculated as $ 2 N \left[ {N - 1 \choose K - 1} - 1 \right] $, while the correct value is
$$
2 K \left[ {N - 1 \choose K - 1} - 1 \right] . \eqno({\rm A}1)
$$

Nevertheless these inconveniences:

\begin{itemize}

\item In the asymptotic expansion of (R18) for $ N \gg 1 $, the contribution $ {\cal O} \left( 1 \right) $ is originated from the {\it tautology} and {\it contradiction} functions.

\item While the wrong reported contribution $ 2 N \left[ {N - 1 \choose K - 1} - 1 \right] $, of the {\it identity} and {\it contradiction} functions, turns out to be $ {\cal O} \left( 1 \right) $, it just adds an extra term $ \ln \left( K_c + 1 \right) $ in (R22) that does not contribute to the $ {\cal O} \left( 1 \right) $ term of its solution (R23). However it gives a wrong, and slower, decaying error $ {\cal O} \left( \ln \ln \ln N / \ln N \right) $.

\item The rest of the Boolean functions, with $ \lambda \geq 2 $, give an $ {\cal O} \left( 1 / N^2 \right) $ contribution to (R18).

\end{itemize}

This implies that all the asymptotic results and their genetical consequences remain correct; while the decaying error term in (R23) becomes $ {\cal O} \left( 1 / N \ln N \right) $ since the correct value (A1) gives a contribution $ {\cal O} \left( 1 / N \right) $ to (R18).

The correct results are obtained as follows:

From (18) and (20), the number of redundances that the elements of $ {\cal I}_K \left( \lambda \right) $ furnish is given by $ \beta_K \left( \lambda \right) \left[ \# \Theta_K^N \left( \lambda \right) - 1 \right] $. From (22), the correct value of $r$ is:
$$
r = \sum_{\lambda=0}^K \beta_K \left( \lambda \right) \left[ {N - \lambda \choose K - \lambda} - 1 \right] . \eqno({\rm A}2)
$$
Note that:

\begin{itemize}

\item The contribution of $ \lambda = 0 $, is the one that corresponds to the {\it tautology} and {\it contradiction} $K$-Boolean functions.

\item The contribution of $ \lambda = 1 $, is the one given by (A1), with $ \beta_K \left( 1 \right) = 2 K $ obtained from (18).

\item The contribution of $ \lambda = K $ is zero. So, irreducible $K$-Boolean functions give raise to injective maps.

\item In the special case of the {\it random map model}~${}^{3,5,13}$: $ r = 0 $ as it should be, due to the fact that, for such a case $ \Psi: {\cal L}_N^N \rightarrow \Lambda_N $ defined by (R4) [respectively by (2) in this article], becomes a bijection so
$$
{\cal L}_N^N \cong \Xi_N \cong {\cal G}_{2^N},
$$
where $ {\cal G}_{2^N} $ is the set of functional graphs from $ 2^N $ points to themselves~${}^{1}$.

\end{itemize}

With this background, the correct equations (R17), (R18), (R19), (R22), (R23), and (R25); are given as follows:

From (R16) and (A2) we obtain
$$
\# \Psi \left( {\cal L}^N_K \right) = \left\{ 2^{2^K} {N \choose K} - \sum_{\lambda=0}^K \beta_K \left( \lambda \right) \left[ {N - \lambda \choose K - \lambda} - 1 \right]  \right\}^N. \eqno({\rm R}17)
$$
Now
$$
\vartheta^{-1} \left( N, K \right) = \left\{ 1 - \varphi \left( N, K
\right)  \right\}^N, \eqno({\rm R}18)
$$
with $ \varphi $ depending also on $N$; and given by
$$
\varphi \left( N, K \right) = {\sum_{\lambda=0}^K \beta_K \left( \lambda \right) \left[ {N - \lambda \choose K - \lambda} - 1 \right] \over 2^{2^K} {N \choose K} }. \eqno({\rm R}19)
$$
From (18), $ \varphi \left( N, K \right) $ admits for $ N \gg 1 $ the asymptotic expansion
$$
\varphi \left( N, K \right) \approx {1 \over 2^{2^K - 1} } \left[ 1 + {\cal
O} \left( {1 \over N} \right) \right];
$$
that gives for the equation $ \vartheta^{-1} (N, K_c) = 1 / 2 $, of the critical connectivity,
$$
2^{2^{K_c}} \approx {2 N \over \ln 2} \left[ 1 + {\cal O} \left( {1 \over N} \right) \right]. \eqno({\rm R}22)
$$
The solution of (R22) is now
$$
K_c \approx \log_2 \log_2 \left( {2 N \over \ln 2} \right) + {\cal
O} \left( {1 \over N \ln N } \right). \eqno({\rm R}23)
$$
And (R25) is now given by
$$
\Delta K_c \approx {2 \over \left( \ln 2 \right)^3 \ \log_2 \left( 2 N / \ln 2 \right) } \sim {\cal O} \left( {1 \over \ln N} \right). \eqno({\rm R}25)
$$

This shows that the asymptotic formulas, and conclusions of Ref.~1 are correct.

\

\bigskip

\vfill \eject

{\bf References}

\begin{itemize}

\item[${}^{1}$] Romero, D., and Zertuche, F., {\it Number of
Different Binary Functions Generated by \textit{NK}-Kauffman
Networks and the Emergence of Genetic Robustness}. J.~Math.~Phys.
{\bf 48} (2007) 083506.

\item[${}^{2}$] Wagner A., {\it Does Evolutionary Plasticity
Evolve?} Evolution {\bf 50} (1996) 1008-1023.

\item[${}^{3}$] Romero, D., and Zertuche, F., {\it The Asymptotic
Number of Attractors in the Random Map Model}.
J.~Phys.~A:~Math.~Gen. {\bf 36} (2003) 3691; {\it Grasping the
Connectivity of Random Functional Graphs}.
Stud. Sci. Math. Hung. {\bf 42} (2005) 1.

\item[${}^{4}$] Kauffman, S.A., {\it Metabolic Stability and
Epigenesis in Randomly Connected Nets}. J.~Theoret.~Biol. {\bf 22}
(1969) 437.

\item[${}^{5}$] Kauffman, S.A., {\it The Origins of Order:
Self-Organization and Selection in Evolution}. Oxford University
Press (1993).

\item[${}^{6}$]Kauffman, S.A. {\it The Large-Scale Structure and
Dynamics of Gene Control Circuits: An Ensemble Approach}.
J.~Theoret.~Biol. {\bf 44} (1974) 167; {\it Developmental Logic
and its Evolution}. BioEssays {\bf 6} (1986) 82; {\it A Framework
to Think about Regulatory Systems}. In: Integrating Scientific
Disciplines. (Ed. W. Bechte) (1986) Martinus Nijhoff, Dordrecht.

\item[${}^{7}$] Wagner, A., {\it Robustness and Evolvability in
Living Systems}. Princeton University Press (2005).

\item[${}^{8}$] de Visser J.A.G.M., {\it et.al.}, {\it
Perspective: Evolution and Detection of Genetic Robustness}.
Evolution {\bf 57} (2003) 1959-1972.

\item[${}^{9}$] Lewin, B., {\it GENES IX}. Jones and Bartlett Pub. (2008).

\item[${}^{10}$] Goebel, M.G. and Petes, T.D., {\it Most of the
Yeast Genomic Sequences are not Essential for Cell Growth and
Division}. Cell {\bf 46} (1986) 983-992; Hutchison, C.A. {\it et.
al.}, {\it Global Transposon Mutagenesis and a Minimal Mycoplasma
Genome}. Science {\bf 286} (1999) 2165-2169; Giaver, G. {\it et.
al.}, {\it Functional profiling of the} \textsf{S. cerevisiae}
{\it genome}. Nature {\bf 418} (2002) 387-391.

\item[${}^{11}$] Thatcher, J.W., Shaw, J.M., and Dickinson, W.J.
{\it Marginal Fitness Contributions of Nonessential Genes in
Yeast}. Proc.~Natl.~Acad.~Sci. USA {\bf 95} (1998) 253-257.

\item[${}^{12}$] Weisbuch, G., {\it Complex Systems Dynamics}.
Addison Wesley, Redwood City, CA (1991); Wolfram, S., {\it
Universality and Complexity in Cellular Automata}. Physica~D {\bf
10} (1984) 1.

\item[${}^{13}$] Derrida, B., and Flyvbjerg, H., {\it The Random
Map Model: a Disordered Model with Deterministic Dynamics}.
J.~Physique {\bf 48} (1987) 971.

\end{itemize}


\

\

\

\

\noindent
Figure caption:

\noindent
Figure~1. (Color online), Graphs for $ {\cal P}_K $  {\it vs.} $ p $ for different values of the average connectivity $ K $. $ K = 1 $ in red, $ K = 1.25 $ in green and $ K = 2 $ in blue. The important $ {\cal P}_K = 1/2 $ value, is in magenta.

\end{document}